\documentclass[mathleft
]{an}
\usepackage{graphicx}
\usepackage{times}
\begin{document}

\Pagespan{288}{}
\Yearpublication{2008}%
\Yearsubmission{2007}%
\Month{11}%
\Volume{329}%
\Issue{3}%
\DOI{10.1002/asna.200710943}

\title{Automated Probabilistic Classification of Transients and Variables}

\author{Ashish Mahabal\inst{1}\fnmsep\thanks{Corresponding author: A. Mahabal,
  \email{aam@astro.caltech.edu}\newline},
S.G. Djorgovski\inst{1},
M. Turmon\inst{2}, J. Jewell\inst{2},
R.R. Williams\inst{1}, A.J. Drake\inst{1}, M.G. Graham\inst{1},
C. Donalek\inst{1}, E. Glikman\inst{1}, \and the Palomar-QUEST Team
}
\titlerunning{Probabilistic Transient Classification}
\authorrunning{A. Mahabal et al.}
\institute{
California Institute of Technology, Pasadena, CA 91125, USA
\and 
Jet Propulsion Laboratory, Pasadena, CA, USA}

\received{2007 Sep 1}
\accepted{2007 Nov 27}
\publonline{2008 Feb 25}


\abstract{%
There is an increasing number of large, digital, synoptic sky surveys,
in which repeated observations are obtained over large areas of the sky
in multiple epochs.  Likewise, there is a growth in the number of
(often automated or robotic) follow-up facilities with varied capabilities
in terms of instruments, depth, cadence, wavelengths, etc., most of which
are geared toward some specific astrophysical phenomenon.  As the number
of detected transient events grows, an automated, probabilistic classification
of the detected variables and transients becomes increasingly important,
so that an optimal use can be made of follow-up facilities, without
unnecessary duplication of effort.  We describe a methodology now under
development for a prototype event classification system; it involves
Bayesian and Machine Learning classifiers, automated incorporation of
feedback from follow-up observations, and discriminated or directed
follow-up requests.  This type of methodology may be essential for 
the massive synoptic sky surveys in the future.
}

\maketitle

\section{Introduction}
Traditional practice of time-domain astronomy generally involves targeted
observing of small samples or even individual instances of a particular type
of variable objects or phenomena.  The recent advent of large, digital
synoptic sky surveys is now revolutionizing the field, thanks to the 
advancement of computing power and detectors (CCDs).  The field has been
moving towards a systematic exploration of larger areas with a better time
sampling and understanding of finer details of many phenomena (e.g., GRBs,
supernovae, variable AGN, etc.).  Many of the events, especially those that
vary on short time scales, need rapid follow-up for proper understanding
and scientific exploitation. This has resulted in a number of robotic
telescopes which can turn to a target very quickly for such follow-ups.

A key link is between event producers (e.g., synoptic surveys, GRB satellites,
etc.) and consumers or follow-up facilities.  The last few years have
seen the emergence of computer networks and protocols which can collect
streams from large surveys and distribute those to facilities that can go
after interesting events. 
The synergy between Palomar-Quest survey
(http://palquest.org) and the VOEventNet system
(http://voeventnet.caltech.edu) 
for the distribution, classification, and follow-up of events
is such an example (Djorgovski et al. 2006, 2007).

In this paper, we describe in more detail the ongoing development of the
automated event classification and follow-up engine for this system.
This experience and methodology should be useful more broadly, for other
synoptic sky surveys, both existing and planned.

As more synoptic sky surveys come online, the problem is going to be one of
plenty.  On the one hand there will be too many events to follow-up 
individually, and on the other hand not all follow-up facilities would be
willing or capable of tracking all types of events due to constraints
on brightness, wavelength, sky visibility, etc.  More importantly, 
most follow-up facilities are generally interested in only specific types
of objects or phenomena (owing to research interests, policy, funding, etc.).
The most critical issue then is of classifying events so that they can be
matched up with facilities ready for them and without unnecessary
duplication of effort.  We note that an automated
event classification of any kind in time-domain astronomy has never been done;
and it may well turn out to be the key enabling technology for the massive
synoptic sky surveys of the future.

Transient event classification is a very challenging problem. A key difficulty 
is the sparsity of information initially available, e.g., position
on the sky and magnitude in one or two bands at a couple of epochs.  
Incorporation of archival data and follow-up observations is essential.
As the available information increases, iterative classification is needed.
One class of existing methodologies is Machine Learning (ML) based.
It includes
Support Vector Machines (SVMs), Artificial Neural Networks (ANNs) etc. 
On the other hand, Bayesian classifiers may be more powerful for these 
applications, owing to the variable and incomplete nature
of the data.  Priors for distributions of observable event parameters
can be formed for different types of objects and
probabilities evaluated for each class.

An important post-classification step is that of feedback based on actual
follow-up, as it will help improve the priors and the resulting classifications.
When classification probabilities are
inconclusive, an intelligent follow-up request engine can suggest the best
follow-up facility to serve as a tie-breaker between two or more competing
classes.

In the next sections we describe the Bayesian classification scheme,
associated supervised learning schemes that can exploit known parameter
dependancies better, revision of priors based on feedback, and the
follow-up request engine.  Throughout the discussion here, transients are
treated as a special class of variables that are typically seen only once
in a given survey (an operational definition), although they may have
counterparts previously seen in other data.
\section{Methodology}
\subsection{Bayesian Event Classification}

Consumers of transient events are usually interested only in particular
kinds of sources, e.g., supernov\ae\ of a given type to be used either as
cosmological standard candles, or as the probes of the endpoints of stellar
evolution; GRB afterglows; gravitational microlensing events, especially
with the possible planetary signatures; flaring AGN; etc.  
Thus the desired output of a classification system is to evaluate a
probability of any given event as belonging to each of the possible known
classes.  Self-imposed probability acceptance cut-off can then allow individual
consumers to decide if a particular event is worth 
following. The most interesting outcome may be the events which do not
fit any of the known patterns and thus are possibly examples of new types of
astronomical objects or phenomena.

Prior distributions need to be estimated for each type of variable astrophysical
phenomena that we want to classify, even though a particular event classification
is inevitably based on incomplete data.
Then an estimated probability of a new event belonging to any given class can be
evaluated from all pieces of information available.
Such information in some format has already been collected by various groups
for particular types of objects, e.g., the Supernova Typing Machine
(http://wise-obs.tau.ac.il/~dovip/typing/)
uses
magnitudes in different filters at different epochs to try and determine the
type of SN a particular event is.  Another example is the search for
quasars in a particular redshift bin based on certain broadband colors.

\begin{figure}
\includegraphics[width=80mm]{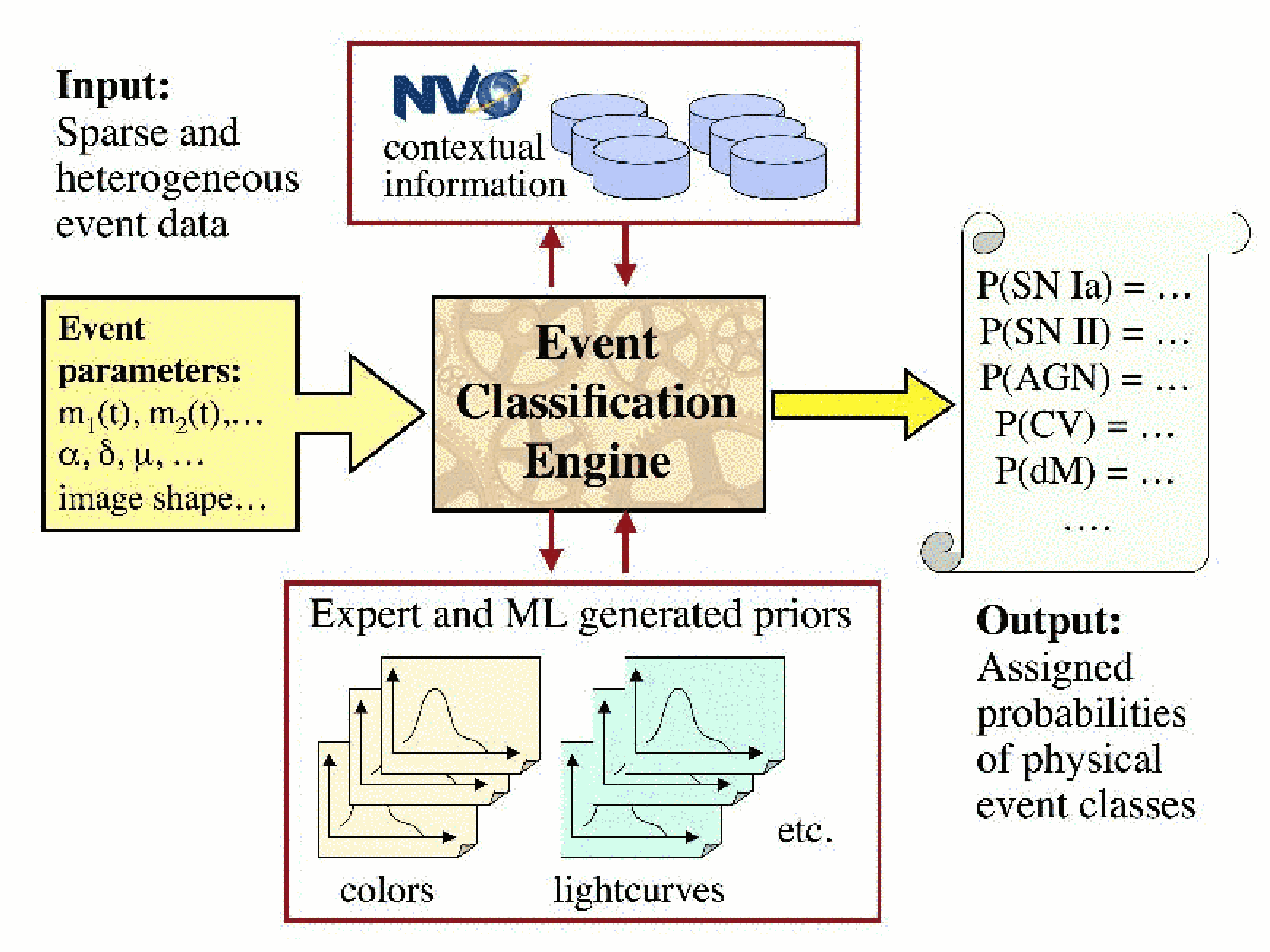}
\caption{
A schematic illustration of the desired functionality of the
Bayesian Event Classification (BEC) engine. The input is
generally sparse discovery data, including brightness in various filters,
possibly the rate of change, position, possible motion, etc., and
measurements from available multi-wavelength archives;
and a library of priors giving probabilities
for observing these particular parameters if the event belongs to a
class X. The output is an evolving set of probabilities of belonging to
various classes of interest.
}
\label{classifier}
\end{figure}

A schematic of the Bayesian Event
Classification (BEC) engine is shown in Fig. \ref{classifier}.
 To take an
example, denote the feature vector of event parameters as $x$, and the object
class that gave rise to this vector as $y$, $1 < y < K$. Many potential entries
within $x$ will be unknown as the information may be incomplete. In practice,
certain fields within $x$ will almost certainly be known, e.g.
sky position, brightness in selected filters etc.. However, other parameters will
be known only selectively: brightness change over various time baselines, and
object shape.  
It is because of the dominance of missing values,
as well as the abundance of prior information that a Bayesian
classification methodology is likely to work best as has been demonstrated by
its effectiveness in such applications as document classification and patient
diagnosis, where there are many sparsely known attributes. 
In this view, $x$ and $y$ are related via
\begin{eqnarray*}
P(y=k|x) &=& P(x|y=k)P(k)/P(x) \\
&\alpha& P(k)P(x|y=k) \\
&\approx& P(k)\Pi_{b=1}^B P(x_b|y=k)
\end{eqnarray*}
Because we are only interested in the above quantity as a function of $k$, we
can drop factors that only depend on $x$. Furthermore, we have assumed
that, conditional on the class $y$, the feature vector decomposes
into $B$ roughly independent blocks, generically labeled $x_b$. These blocks may
be singleton variables, or contain multiple variables - for example, sets of
highly correlated filters.
The decoupling allows us
to (1) circumvent the curse of dimensionality as the decomposition keeps the
dimensionality of each
block manageable (we will eventually have to
learn the conditional distributions $P(x_b | y = k)$ for each $k$. As more
components are added to $x_b$, more examples will be needed to learn the
corresponding distribution),
and (2) cope easily with
ignorance of missing variables by dropping the corresponding factors from
the product above. 
The methodology makes a seemingly strong independence
assumption, but in practice, because we are after a classification and not an
exact membership probability, classification results can be excellent even
when the assumption is violated (Hand \& Yu, 2001; Domingos \& Pazzani 1997).
\subsection{Machine Learning Event Classifier}
Besides the 
BEC,
traditional supervised classification
methods such as the ANNs and/or 
SVMs (Vapnik 1995; Cristianini \& Shawe-Taylor 2000; Fan,
Chen \& Lin 2005; Ripley 1996; Scholkopf \& Smola 2001)
can be used for confirmed event databases with large,
nonsparse training and validation data sets where the use of supervised
networks is already well established. Such a Machine Learning Event
Classifier (MLEC) can represent events as vectors of observable parameters
$X = {x_1, ...,  x_i,  ..., x_n},$
where $x_i$ are
various
observed quantities for
the large majority of events, e.g., flux amplitudes in various filters,
coordinates, flux ratios, etc.

\begin{figure}
\includegraphics[width=80mm]{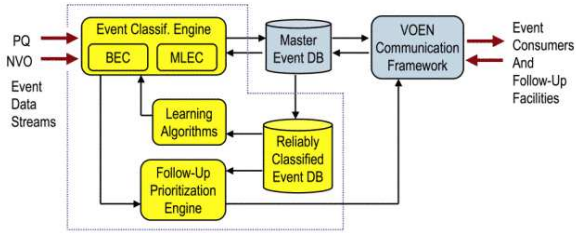}
\caption{
The Event Classification Engine, containing two separate classifiers described
in the text (BEC and MLEC) would provide event classifications, and incorporate
newly obtained data from follow-up observations for improved, iterative event
classifications.
Reliably classified events
can be used for
refinements of the event classifiers, as well as for the operation of the Follow-up
Prioritization Engine (FPE).
}
\label{classifier2}
\end{figure}

Two types of problems can be expected:
(1) not all parameters
would be measured for all events. For example, some may be missing a
measurement in a particular filter, due to a detector problem; some may be in
the area on the sky where there are no useful radio observations; etc. 
A partial solution is to train a set of quasi-independent classifiers
and invoke the one most suited based on observations available.
(2) many observables would be given
as upper or lower limits, rather than as well defined measurements.
This can be partly solved by treating them as actual measurements or missing
values leading to inaccurate or lossy data.
Thus, this approach may be more useful for a
classification of variable (always present, but changing) sources, rather than
transients (detected only once). However, the performance of MLEC
would be constantly improving as more follow-ups happen.
A schematic combining the
BEC, MLEC and the feedback stages is shown in Fig. \ref{classifier2}.
\subsection{Feedback Incorporation}
A crucial feature of the system should be the ability to update and revise the
prior distributions on the basis of the actual performance, as we accumulate
the true physical classifications of events, e.g., on the basis of follow-up
spectroscopy.  Learning, in the Bayesian view, is precisely the action of
determining the probability models above - once determined, the overall
model (1) can be used to answer many relevant questions about the events.
Analytically, we formulate this as determining unknown distributional
parameters $\theta$ in parameterized versions of the conditional distributions
above,
$P(x | y = k; \theta)$. (Of course, the parameters depend on the object
class $k$, but we suppress this below.) In a histogram representation, $\theta$ is
just the probabilities associated with each bin, which may be determined by
computing the histogram itself. In a Gaussian representation, $\theta$ would be
the mean vector $\mu$ and covariance matrix $\Sigma$ of a multivariate Gaussian
distribution, and the parameter estimates are just the corresponding mean and
covariance of the object-$k$ data.  When enough data is available we can adopt a
semiparametric representation in which the distribution is a linear
superposition of such Gaussian distributions:
$$P(x_d|y=k;\theta)=\displaystyle\sum_{m=1}^{M} \lambda_m N (x_d;\mu_m;\Sigma_m)$$
This generalizes the Gaussian
representation, since by increasing $M$, more distributional characteristics may
be accounted for. The corresponding parameters may be chosen by the
Expectation-Maximization algorithm (Turmon, Pap \& Mukhtar 2002)
or kernel density estimation
(Silverman 1986; John \& Langley 1995).
Three possible sources of information
can be used to find the unknown parameters: (1) background physical
knowledge, e.g. from
considerations of monotonicity,
(2) examples
labeled by experts, (3) feedback from the downstream observatories
once labels are determined. The first case gives an analyical
form for the distribution, but the last two provide
labeled examples, $(x, y)$, which can be used to select a set of $k$ probability
distributions as described above. 
The parallel performance of the Bayesian and Machine
Learning event classification engines can be evaluated and compared,
and the output of both used - unless
one turns out to be clearly superior to the other. 
\subsection{Follow-up Prioritization Engine}
\begin{figure}
\includegraphics[width=80mm]{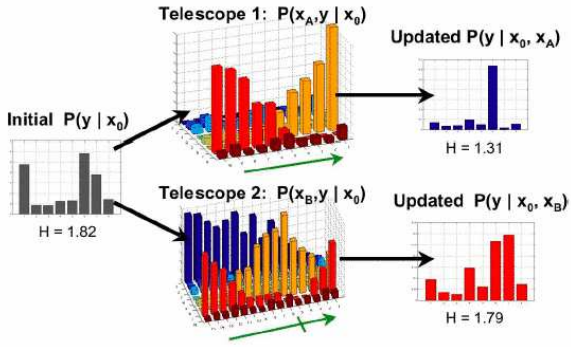}
\caption{
A schematic illustration of follow-up observation recommendations:
At left, the initial estimated per-class
probabilities for eight object classes, showing high entropy resulting from
ambiguity between the object classes numbered 1, 6, and 7. 
Followup observations from two telescopes are possible (center). Their resolving
capacity is shown as a function of class $y$ (left axis) and observed value
(right axis parallel to green arrows). In the diagram, for telescope 1, as
observed value $x_A$ moves up the green arrow, class 6 becomes increasingly
preferred. For telescope 2, moderate values (near the crossbar in the arrow)
indicate class 6, and other values indicate class 7. Finally, at right,
typical updated classifications. The lower-entropy classification at the top
is preferred. Since the particular values used for refinement ($x_A, x_B$) are
unknown at decision time, appropriate averages of entropy must be used, as
described in the text.
}
\label{discriminatoryobs}
\end{figure}

The sparse data can often lead to cases of ambiguous classification
or perhaps it may not lead to
meaningful classifications at all. On such occasions a
follow-up prioritization engine can suggest the best follow-up strategy to
reduce confusion between competing classes.
For example, the
system may decide that obtaining optical light curve with a particular time
cadence would discriminate between a Supernova and a quasar, or that a
particular color measurement would discriminate between a cataclysmic variable
eruption and a gravitational microlensing event, etc. Suitable prioritized
requests for the needed follow-up observations would be generated and sent to
the appropriate telescopes.  Since 
observational resources are scarce it is important to
rank order the possible follow-up observations according to which ones result
in the most reduction in classification uncertainty. This can be done using an
information-theoretic approach 
(Loredo \& Chernoff 2003)
by quantifying the classification uncertainty using the conditional
entropy of the posterior for $y$, given all the available data. When an
additional observation, $x_+$, is taken, the entropy decreases from
$H(y|x_0)$ to $H(y|x_0, x_+)$.
This is illustrated in Fig. \ref{discriminatoryobs},
where the original classification
$p(y | x_0)$
is ambiguous and may be refined in one of two ways. The refinement for
particular observations $x_A$ versus $x_B$ is shown.
The correct choice is the one that will reduce
the final entropy most.
In our notation, the best follow-up observation is the one which results in
the minimal final entropy, given by
\begin{eqnarray*}
X_+ &=& min_{x_+} H(y| x_+,x_0) \\
&=& -\displaystyle\sum_{yx_+}
p(y,x_+|x_0) \ \rm{log}\  p(y|x_+,x_0)
\end{eqnarray*}
In computing this, we average over all possible values of the new measurement
$x_+$ and class $y$.  Note, this is equivalent to maximizing the conditional mutual
information of $x_+$ about $y$, given $x_0$; that is, 
$I(y; x_+ | x_0)$
(Cover \& Thomas 1991).
The joint
density above is known within the context of our assumed statistical model.
Specifically, we have a joint probability of the form:
$$p(y,x_+|x_0) = {p(x_+,x_0|y)p(y) \over \displaystyle\sum_{yx_+}
p(x_+,x_0|y)p(y)}$$
\noindent where the right hand side is given by factors as in (1). The conditional
probability,
$$p(y| x_+,x_0) = {p(x_+,x_0|y)p(y) \over \displaystyle\sum_y
p(x_+,x_0|y)p(y)}$$
\noindent is the Bayes posterior given the new and previous measurements. Therefore, we
can compute, within the context of the previously learned statistical model
used to define the posterior in (1), the follow-up measurement resulting
in the greatest entropy reduction given the previous measurements. We can thus
provide a rank-ordered list of potential follow-up observations according to
the information-theoretic ranking, leading to the most efficient use
of the resources.
\section{Summary}
We have presented a software methodology, now under development,
for an automated, iterative probabilistic classification of variables
and transients found in large digital synoptic sky surveys.
Our primary approach is using Bayesian networks, with a parallel
development of classifiers based on Machine Learning techniques.
Incorporation of feedback from follow-up observations is essential, both
to update the Bayesian priors, and to improve the training data sets for
the ML algorithms.  Another innovation is  
an engine to discriminate between possible follow-up facilities for
optimal results as well as faster rate of learning.  Experimental
implementations of this methodology with existing surveys such as PQ
should both enhance their scientific returns, and help lay the groundwork
for the more ambitious projects in the future, such as PanSTARRS and LSST.
\acknowledgements
We are grateful to the members of the PQ survey team, and to the staff of
Palomar Observatory.  This work was supported in part by the NSF grants
AST-0407448, AST-0326524, and CNS-0540369, and by the Ajax Foundation.  SGD acknowledges a stimulating atmosphere of the Aspen Center
for Physics.  Finally, we thank the workshop organizers for an excellent and
productive meeting.

\end{document}